# INFLATION AND VALUE CREATION: AN ECONOMIC AND PHILOSOPHIC INVESTIGATION

Gennady Shkliarevsky

**Abstract**: The subject of this study is inflation—a problem that has plagued America and the world over the last several decades. Despite a rich trove of scholarly studies and a wide range of tools developed to deal with inflation, we are nowhere near a solution of this problem. We are now in the middle of the inflation that threatens to become a stagflation or even a full recession; and we have no idea what to prevent this outcome. This investigation explores the real source of inflation. Tracing the problem of inflation to production, it finds that inflation is not a phenomenon intrinsic to economy; rather, it is a result of inefficiencies and waste in our economy. The investigation leads to a conclusion that the solution of the problem of inflation is in achieving full efficiency in production. Our economic production is a result of the evolution that is propelled by the process of creation. In order to end economic inefficiencies, we should model our economic practice on the process that preceded production and has led to its emergence. In addition, the study will outline ways in which our economic theory and practice must be changed to achieve full efficiency of our production. Finally, the study provides a critical overview of the current theories of inflation and remedies that are proposed to deal with it.

## Introduction

Inflation is a very familiar word in economics. It is also a highly popular topic these days. Not a day passes without hearing this word from politicians, economists, mainstream media sources, or even in casual conversations. The reason for this popularity is the fact that the world economy is in the grips of inflation, and not just an ordinary one. For one thing, the inflation is high. Even in the U.S. where the level of inflation is now the lowest in the world, it is close to 9%. This inflation does not affect one particular or even group of countries; it is a truly worldwide phenomenon. Even in developed G-7 countries it reaches over 10% and is even much higher in less developed and underdeveloped countries.[1] It is a global phenomenon. Fabio Vighi offers this characterization:

> We have entered a global cycle of secular inflation that is unique in
> history. The cynical attempt to preserve a system based on the ontological assumption of permanent monetary injections now entails the controlled demolition of the real economy and the world it supports.[2]

So far, all efforts to curb this inflation by raising interests rates have failed. Many observers point to an imminent danger of this inflation becoming a run-away inflation or hyperinflation. Even worse, it can turn into a stagflation, similar to the one that erupted in the mid-1970s that had very severe economic and political ramifications from which our economy has not recovered to this day.

The consequences of the current inflation can be even worse. For example, if the inflation of the 1970s led to the demise of the Democratic Party for over a decade, this inflation may very well wipe it out completely. The current inflation has already led to the wave of political discontent. This wave has already led to the election of Donald Trump and to the establishment of Republican control over the House of Representatives. There is no doubt that this inflation will play a very important role in the 2024 elections. Many researchers fear that this inflation will bring upon us increasing "ideological manipulation and authoritarian violence."[3]

Inflation has always been one of the macroeconomic phenomena that attracted much attention. Inflation and its counterpart, deflation, have always provoked heated debates.[4] Decades of research have expanded our knowledge of various factors that drive inflations. Although we have come to understand a lot about inflation, we still do not have a clear idea as to the root cause of inflation. We are not even sure whether inflation is necessarily a bad thing.[5] In fact, many economic planners currently believe that we need inflation and one or two per cent inflation is actually good for our economy. We have also learned that deflation--the opposite of inflation--is not much better than its counterpart. Japan is a sad example of its pernicious impact.[6] The result of the confusion and controversies surrounding inflation is the lack of coherent and clear policies that would curb the current inflation. Despite the fact that the Fed is currently increasing interest rates to fight the inflation, there are many prominent economists who criticize the Fed's approach. Joseph Stiglitz sees in it nothing but pain and no benefits. He writes: "As a new Roosevelt Institute report that I co-authored shows, any benefits from the extra interest-rate-driven reduction in inflation will be minimal, compared to what would have happened anyway."[7] We seem to be permanently deadlocked as we are trying, on one hand, to avoid high unemployment and, on the other, to prevent run-away price increases.

The purpose of this article is to find a way out of the current conundrum related to inflation. As part of the search for a solution, the article will provide a critical survey of the current views on inflation and popular approaches in dealing with it. Much of the current research on inflation focuses on various factors that drive inflations. No doubt, this knowledge is important but it is not sufficient. The factors that drive inflation are relevant but they seem to be phenomena that attend inflation, rather than cause it. For example, we certainly accept the notion that growing money supply drives price increases, but this notion tells us nothing about the reason why our policy planners decide to boost the supply of money, even though we know perfectly well the consequences of such policy. In many cases, economic planners resort to this policy in order to contain unemployment, which suggests that our economy is always in a double jeopardy, oscillating between inflation and unemployment. There is a vital need to determine what and why sets price stability and employment on a collision course.  This article will address all these questions and issues related to them.

## Current Views on Inflation and Its Causes

### *Definitions*

The term "inflation" in relation to economic processes came into use in the second half of the 19[th] century.[8]  Back then the term "inflation" referred primarily to the expansion of the currency.  This view is still popular today, particularly among neo-classical economists.  According to Milton Friedman, for example, inflation "is always and everywhere a monetary phenomenon."[9]  Only later, economists began to define inflation in terms of price increases that might result from monetary expansion but could also have other causes.[10]

Today, most economists think about inflation as price increases for goods and services, or "how much more expensive the relevant set of goods and/or services has become over a certain period, most commonly a year."[11]  There are certainly wordier definitions.  *The Encyclopedia Britannica* offers the following expanded version:

> Inflation, in economics, is collective increases in the supply of money, in money incomes, or in prices.  Inflation is generally thought of as an inordinate rise in the general level of prices.[12]

In his recent contribution, Adam Tooze offers an even lengthier formulation:

> In discussing inflation, economics abstracts from idiosyncratic shocks. Inflation is defined as a general upward pressure on all prices, independent of idiosyncratic supply shocks. Inflation, in this sense, is a macroeconomic, aggregate concept . . . Inflation can thus be defined as a shift in the terms of trade between (1) money and (2) goods, as experienced (3) by a particular group of people and (4) captured by a particular statistical apparatus.[13]

No matter how long or how sophisticated current definitions are, none of them really ventures beyond the mere description of inflation and factors attending to it.

### *Perceived Causes of Inflation*

There is no shortage of theories that try to explain what causes inflation.[14]  They are different and diverse.  Yet all they do is create confusion.  They all focus on some specific factor or factors attending inflation and try to reduce their explanation what they see as the cause.  The factors they stress are often valid, but together they create a contradictory picture that defies understanding.

Perhaps the most common explanation of what causes inflation is a monetarist one. It is also by far the oldest. David Hume recognized the link between the expansion of money supply and price increases in his essay "Of Money" that appeared in 1752. This explanation remains popular to this day.[15] According to this school of thought, there is only one cause of inflation and it is an over-issue of inconvertible money.[16] Milton Friedman is the best-known representative of the monetarist school. In his concise, if restrictive, formulation, inflation is a monetary phenomenon. As he put it, inflation "is and can be produced *only* by a more rapid increase in the quantity of money than in output."[17]

The reason for the continued popularity of the monetarist explanation is the fact that it enjoys more empirical support than any other economic theory.[18] However, despite its long history and the substantial factual evidence, the predicted association between money supply and inflation remains disputed.[19] For one thing, this association works only over long periods of time. Also, there are some examples of variations in money supply that do not necessarily negatively correlate with price levels. For example, decreases in the supply of money do not necessarily lead to price increases. During the Great Depression decreases in money relative to real incomes were associated with decreases, not increases, in price levels.[20]

Also, the monetarist explanation of what causes inflation is not unambiguous. The link between money supply and the availability of goods/values goes both ways. If the increase in money supply can cause price increases, so can the decline in the production of goods. The connection between inflation and production is a common preoccupation among economists. David McMillan, for example, explains the high rate of inflation in Britain by the sluggish growth of productivity and low rate of investment that lags behind Britain's main economic competitors—the United States, Germany, and France.[21] The results of a study that has tested the link between inflation and economic growth in Sri Lanka show that there is "a long run negative and significant relationship between the economic growth and inflation."[22] The study shows a similar but short-term impact for China.[23] A study by Hrushikesh Mallick also reveals the negative correlation between inflation and growth in India.[24]

Government spending is yet another popular explanation for inflation. Conservatives have expertly used this explanation in the past as a way of defeating their opponents who favor the expansion of government programs. Ronald Reagan effectively employed this cudgel by arguing that extensive and expensive government programs had a devastating effect on American economy and were the principal cause of the stagflation that occurred in the mid-1970s. One can hear very similar criticisms of the government these days. In his contribution "Inflation and Growth: Some Theory and Evidence" Max Gillman challenges liberal economists and politicians who explain the current high rate of inflation by the war in Ukraine, "corporate greed," or short-term supply issues. The cause, in his view, is "bad public policy."[25] Nathan Benefield makes a similar charge in his editorial. "Washington politicians," he opines, "pretend they don't know what's behind runaway inflation, but voters know." Pointing to what he sees is the real issue, Benefield quotes James Carville, a Democratic strategist, who quipped on one occasion: "It's the spending, stupid."[26]

This list of explanations is far from complete. The number of factors that researchers cite as causing, contributing to, or driving inflation is much larger. It is beyond the scope of this study to examine them all. They include supply bottlenecks, unequal distribution of money, sluggish price adjustment, too "few" unemployed, and many others.[27] There are even sociological perspectives on inflation that focus on the role, interests, and expectations of social actors as a cause of inflationary tendencies.[28] In reading various authors, one gets an impression that researchers base their explanations on their theoretical or political commitments; and that what they present as causes of inflation are merely their subjective preferences among factors that attend inflations. Gregory Mankiw appears to say as much when he writes about economists "dirty little secret" since they often speak "not just as economic scientists, but also as political philosophers."[29] Alexander Barta observes:

> . . . [I]t is elementary to recognise that the very measurement of inflation, i.e. the determination of the basket of goods and services whose prices are taken into account as much as the very production of the statistic itself, is a selective, theory-driven and, indeed, political choice already--there is no such thing as "objective inflation" . . . Data are business. Data are political. And that is particularly pertinent in the case of inflation, because inflations are contentious. They generate winners and losers.[30]

There are no known factors cited as a cause of inflation that would not be put in doubt by empirical evidence. Even the most commonly cited cause—money growth—is not without counter examples. As has already been mentions, during the Great Depression of the 1930s money relative to income and the price levels were both in decline.[31]

The examination of the current explanations of what causes inflation shows that researchers have different views on this subject. The causes they cite—supply of money, production rates, structure of the economy, distribution of wealth and money, savings and shopping habits, and many others—are merely co-dependent economic variables attending inflations, rather than causing it. Barta correctly points out that inflation cannot follow from any of these co-dependent variables because variations in any of them simply leads to "recalibration of the system of relative prices by the Walrasian Auctioneer, but not to a rise in the general price level."[32] On close analysis, what is often perceived as a cause of inflation seems to be a symptom—a variable attending to inflation—rather than its cause.

*Inflation Remedies*

There are two principal approaches in dealing with inflation. One emphasizes the role of the government in managing inflation. The other recommends relying on market forces, rather than government fiat, to achieve price stability.

    1. Inflation Management

The roots of the first approach rest on Keynesian economic theory. Keynes challenged neoclassical economics and its reliance on market forces to regulate the economy. The pro-active approach advocated by Keynes proposed that the government was to use a variety of levers available to it to manage economy and inflation.

According to this approach, inflation is neither bad, nor good; or rather, it is bad and good at the same time. Good inflation is one that stays at a recommended level of 1% to 2% a year or 10% to 12% for developing countries.[33] The proponents of this approach maintain that while high inflation can be harmful, too little inflation or even negative inflation, or deflation, can also weaken the economy. Japan, for example, has experienced a long period of practically no economic growth primarily because of deflation. Following the financial crisis of 2007, the Fed and other central banks around the world promoted low interest rates and other monetary incentives to make sure that liquidity stays sufficient; they euphemistically called it "quantitative easing."[34]

The main thrust of the Keynesian approach and its variations is to provide a framework in which the government can manage economic activity by varying its expenditures and receipts or by influencing the level of private investment through interest rates and money supply.[35] The followers of Keynes do not seek to eliminate inflation. Rather they try to ameliorate its most harmful effects and keep it at a manageable level.

The Federal Reserve, or the Fed, is the main institution that regulates the economy. The Fed has two jobs: one is to maintain price stability and the other is to maximize employment. When the economy is struggling and inflation is too low, the Fed will lower interest rates or buy assets to increase the amount of available cash. When the economy is expanding too quickly and inflation rises, the Fed will typically raise interest rates or sell assets to reduce the volume of cash in circulation. Borrowing money becomes expensive, which slows economic growth and brings down the level of inflation.[36] The theory that guides the Fed and most central banks aims at mild inflation at a level of one or two per cent.[37] However, this theory is little more than just a theory.

The policy actions of the Fed and other central banks in the face of inflation may include raising interest rates at which central banks provide reserves to financial intermediaries, buying and selling government securities, changing reserve requirements, and others.[38] This approach is very eclectic. Its proponents offer a variety of policies and policy combinations that are supposed to help curb inflation. They propose to lower health costs, reform the tax code to raise more revenue, limit discretionary spending, reduce consumption-oriented spending, cut aid to states, reduce costs of energy, trade, and procurement, and much else.[39] Even a more restrained fiscal policy is not off-limits. As President Biden has remarked numerous times in the current inflation period, "bringing down the deficit is one way to ease inflationary pressures." This could include avoiding further deficit-boosting measures; lowering health care costs; raising tax revenue; reducing consumption-oriented spending; promoting work, savings, and investment; and/or lowering energy, trade, and procurement costs.[40]

Even when the American government abandoned the policy of low interest rates in response to inflationary pressures, many Keynesians criticized this move by insisting that giving up this policy was wrong. Joseph Stiglitz, for example, has argued in his article that in a recent issue of *Project Syndicate* that high interest rates bring nothing but pain to the economy and have no benefits.[41]

One should not get an impression that this approach to inflation is limited to Democrats. President Donald Trump throughout his presidency publicly pressured the Federal Reserve to keep interest rates low and to keep pursuing bond-buying monetary expansions. In one of his tweets, Trump wrote in April 2019:

> We have the potential to go up like a rocket if we did some lowering of rates, like one point, and some quantitative easing . . . Yes, we are doing very well at 3.2% GDP, but with our wonderfully low inflation, we could be setting major records &, at the same time, make our National Debt start to look small!"[42]

As the current inflation continues to grow, even the current American government that is dominated by Keynesians has launched a policy of high interests in expectation that this measure will reduce inflation even if at the cost of employment. One can surmise that the intention now is to keep on alternating between the policy of high interest rates and the policy of high employment. There is not much theory behind this constant alternation; it seems to be more of a trial-and-error approach driven by wishful hope with little empirical support that somehow the economy will grow out of the current predicament.

## 2. Price control

Perhaps the most radical policy that exemplifies the approach that favors government intervention is price control. There are few economists and politicians who advocate this policy, but it is on the menu of possibilities. One can never trust politicians who may be opposed to price control one day and then change their position the following day.

Power brokers tried this policy in the 1970s when inflation was devastating family earnings and when the buying power of common Americans was severely reduced.[43] President Nixon preached to Americans in 1965 that "the lesson that government price fixing doesn't work is never learned." When campaigning for president, Nixon pledged that he would "not take this nation down the road of wage and price controls." However, in 1971, when inflation reached six per cent, Nixon, against his own previous judgment, began to pressure the business community, with little or no success, to hold down prices and wages. Prices continued to grow.[44] Trying to deflect criticism, President Carter blamed rising prices and recession on OPEC, also with little success.

President Biden's initial response to rising prices was dismissive. He simply declared them temporary. When prices continued to grow, Biden blamed rising prices on the pandemic. When the war in Ukraine erupted, Biden quickly put the blame on Putin. In his view and that of other Democrats, the rising prices justified emergency action. Elizabeth Warren and the host of other left-learning Democrats rushed to vilify "greedy corporations" for exploiting the pandemic. Later in 2021 *The New York Times* ran a trial balloon. The paper wrote approvingly that as rising inflation was threatening Biden's presidency, he was turning to "the federal government's antitrust authorities to try to tame red-hot price increases."[45] Politicians certainly took the cue. Representative Jamaal

Bowman (D-N.Y.) obliged the president by introducing the Emergency Price Stabilization Act, calling on the government to "build the capacity to establish limits on the growth of certain prices, and to otherwise strategically regulate such prices, in order to stabilize the cost of essential goods and services."[46] Targeting price stability as a way to achieve higher rates of economic growth is also popular in countries with emerging markets.[47]

### 3. The Neo-Classical Approach

As has already been mentioned, the other approach toward inflation has originated in neo-classical economics. The advocates of this approach stress that the availability of resources to produce goods, services, and particularly technological change is a major factor affecting growth.[48] The theory that underlies this pro-growth perspective maintains that producing more goods and services in a shorter time would cut costs per unit, raise supply, and thus put downward pressure on prices.[49] The secret to curing inflation, in their view, is pro-growth policies that create incentives for more goods, more employment, less government spending and sound money. When production grows, prices go down.[50] However, in contrast to interventionists, the pro-market proponents insist on reducing the role of the government.

The policies pursued in the market-oriented approach stress the need to reduce government interventions into the economy. Such reduction takes several forms. First of all, it involves severe restrictions to the government's capacity to increase the supply of money—either by limiting the issue of paper money or by preventing the government from manipulating interest rates. The idea is that if the government stops playing with money supply, investors can get a realistic picture of their opportunities, better calculate risks, and have sober expectations about profits.

Also, the proponents of the market-oriented approach advocate the reduction of the federal budget by eliminating high-cost government programs. Cutting government programs was a signature policy introduced by Ronald Reagan and continued under Bill Clinton. Lawrence Summers, who served as Bill Clinton's Treasury secretary, rocked the Democratic establishment last year by predicting that his party's excessive spending would cause inflation.[51] For Nathan Benefield, Senior Vice President of the Commonwealth Foundation--Pennsylvania's free-market think tank—government spending is the main factor that generates inflation. He succinctly summarizes the reason for the current inflation by paraphrasing the Democratic strategist James Carville: "It's the spending, stupid." In his view, if policy makers are serious about fighting inflation, they should start with fiscal restraint.[52]

In accordance with the pro-market approach, reduction of government spending should go hand-in-hand with cutting taxes. Tax cuts will give more money to both business owners and regular working families; and more money for them will invigorate the market, ensure growth, and cut inflation. There is no shortage of calls from both the right and the left on lawmakers demanding to enact a comprehensive tax reform that would ease burdens on working families and enable small businesses to hire more workers and raise wages. In their article "Economic Growth, Not Austerity, Is the Answer to Inflation" Arthur Laffer and Stephen Moore cite historical examples that prove their point. They write:

> History proves growth doesn't cause inflation. In the 1920s, when the highest tax rate was cut from 73% to 25%, real GDP soared and the price level fell. In the 1960s, tax cuts and pro-growth policies led to an economic expansion, stable prices and budget surpluses.[53]

Lastly, many economists and politicians also see the need to reform or even eliminate various excessive regulations imposed by the government that constrain economic activity and negatively impact production. Those who support reducing regulations argue that such reforms would help producers to control costs and unleash energy production to lower electricity and fuel prices.[54]

*Critique of the Proposed Remedies*

As the above shows, recipes for dealing with inflation are different and diverse. The proponents of each approach mostly focus on the advantages of what they propose and prefer to ignore disadvantages. However, an objective assessment needs to take into account all sides of what is proposed—both positive and negative. Criticisms in this section will focus on shortcomings of these recipes, both those specific to each one and those common to all.

1. Managing Inflation

Perhaps the most serious disadvantage of the managing inflation approach is the fact that it has proven to be completely ineffective in the current inflation period. The proponents of this approach have claimed that maintaining a tolerable level of inflation at one or two percent a year is relatively harmless and will have no serious ramifications for the economy. In their view, this approach should allow making financially sound decisions on saving, investing and borrowing money.[55]

However, trying to manage inflation at a desired level has proven to be very difficult. Since 1921 secular inflation in the United States and throughout the world has been a permanent presence in the economy. There have been rises and declines. The current secular inflation has grown steadily in the last decade or so and reaches today the level of close to 9% in America. Among all developed countries, the inflation is the highest in Great Britain reaching the ominous 10%. All through this period, the Democrats defended and continue to defend the policy of high government spending and expensive government programs. The evidence from both industrialized and developing countries supports the view that inflation causes lower real growth. As Woo S. Jung and Peyton Marshall conclude in their article on inflation and economic growth: "The use of inflationary finance as a means to force additional savings and to increase capital formation appears to be an unwise strategy for economic development."[56] Manipulating different variables to maintain the recommended level of 1% to 2% inflation is like balancing economy on a pinhead: even if you achieve this balance, it will be extremely precarious, affected by various destabilizing economic and non-economic factors.

In extreme cases inflation can become a run-away inflation, or even hyperinflation.[57] One of the best-known cases of hyperinflation is Weimar Germany in 1923 when one American dollar was worth 4 trillion German marks. Money lost all value and people shifted to barter, using goods with stable value as currency. For example, pianos became currency during the German inflation.[58] The post-war inflation in Hungary reached 41.9 quadrillion percent.[59] In 2008, Zimbabwe experienced one of the worst cases of hyperinflation ever. The estimated annual inflation level at one point was 500 billion percent.[60]

High inflation can cause a decline of production. When high inflation and production decline occur at the same time, the result is what is called stagflation. Stagflations have particularly devastating effects. One may recall the stagflation in America during the mid-1970s when high interest rates, rising prices, and economic stagnation created an extremely explosive political situation that led to the collapse of the Democratic Party and brought Ronald Reagan to power and almost a decade of Republican rule.

The current inflation also has a very strong potential to turn into a stagflation.[61] Top economists and bankers have already cautioned about this possibility. Those who have sounded alarm include Allianz and Gramercy's chief economic adviser Mohamed El-Erian and Goldman Sachs CEO David Solomon. Both have pointed out that the current inflation is entrenched and widespread throughout the world. The World Bank has also issued multiple warnings to the effect that if economy remains sluggish, inflation may very well end in a stagflation in many countries, including in the United States.[62] Even Paul Krugman, of *The New York Times*, who is usually optimistic regarding current inflation, has expressed concerns over the persistently "hot job market."[63] One economist writes about the current mood among economists that "they also know that this hardly original escamotage can only lead to runaway inflation, and then hyperinflation. What takes place today as a matter of monetary normality used to characterise wartime economies, namely direct financing via the money presses."[64]

Many leading economists fear that today's high levels of inflation, and the Fed's commitment to containing it, could trigger a recession as early as 2023.[65] For example, the vast majority of economists at 23 large financial institutions have already confirmed this gloomy prediction only a few days into 2023.[66] In his article for *Project Syndicate*, Nouriel Roubini predicts an "unavoidable crash" of the economy in the near future.[67] The Fed is currently trying to contain inflation with high interest rates that should eventually reach 5.1%. At this rate, the unemployment may reach the level as high as 4.6% and possibly even higher. Yet despite these efforts the inflation remains resilient. Even economists who try to moderate fears of recession, such as Jeffrey Frankel, still acknowledge that in the next two years a worldwide recession is entirely within the range of possibilities.[68]

2. Price control

Price control is a very radical policy that should be used with great caution. It is certainly a legitimate policy that has often been used in dire circumstance of war when goods are in short supply. Governments successfully used this policy in combination with the policy of distribution of resources. However, using this policy to fight inflation is totally useless and even damaging.

The government controlled prices in the Soviet Union through much of the country's history. It worked reasonably well in the time of WWII when many goods, particularly food products, were very scarce. However, continuing this policy in peacetime, particularly in the late 1970s and early 1980s, resulted in horrible economic distortions and inefficiencies. Indeed, the inflation was low on Soviet planners' books, but it was merely pushed underground and out of sight. Money continued to lose their value but prices were kept stable by government fiat. Many citizens began to run away from cheap money into goods with stable and low prices. The most notorious example of this flight was the fact that many farmers used cheap bread to feed their livestock. The result was enormous hoarding and chaotic disappearance of goods. By not addressing the source of inflation and pretending that it did not exist, just because prices did not change, Soviet economic planners simply made inflation uncontrollable.

There are many stories about that period in the late 1970s and 1980a in the Soviet Union. Many who observed the erstwhile Soviet economy documented empty shelves and long lines to buy even the most essential items (toilet paper and vodka were two items particularly in high demand). Researchers habitually attributed these shortages to production inefficiencies. Indeed, production in the Soviet Union was inefficient, but no more than usually. The real story behind these shortages was that farms and industries continued to produce goods at their normal rate, but these goods disappeared into the thriving black market even before they left enterprises. Crafty speculators hoarded goods and sold them for high profit. Many goods, particularly perishable food products, rotted in underground warehouses, while people waited for days in huge lines to buy even the semi-spoiled leftovers.

Such are the lessons of price control. No wonder that political leaders rarely resort to this policy, despite occasional calls from academics.[69] President Nixon, for example, who at one point talked about price control, still relied on appeals to the business community to limit wages and prices on a voluntary basis.

3. Reducing the Role of the Government in the Economy

The approach that sees the solution of the problem of inflation in reducing the role of government in the economy and in relying on market forces to restore economic stability has several major shortcomings. For one thing, the reduction of money supply as a cure for inflation will inevitably slow down production, which will also inadvertently lead to unemployment. The proponents of this approach argue that high level of unemployment is simply unavoidable. It is, in their view, a necessary evil that we simply have to endure. However, many people who end up on the receiving end of this policy and have to pay a heavy price for it are not particularly receptive to the idea that their suffering paves the road to future happiness. Unemployment, declining living standards, growing gap between the rich and the poor are sure to lead to political tensions, ravaging instability, and social cataclysms.

Many neo-classical economists advocate this perspective.  Ronald Reagan was perhaps its best-known practitioner.  However, there have also been quite a few Keynesians among those who recommended this approach.  In 1980, for example, when inflation in America stood at double-digit, Paul Samuelson, who was the first American to win the Nobel Prize in economic sciences, wrote that "five to ten years of austerity, in which the unemployment rate rises to an eight or nine percent average and real output inches upward at barely one or two percent per year, might accomplish a gradual taming of U.S. inflation."[70]  Closer to our time Lawrence Summers, who served as Bill Clinton's Treasury secretary, has also argued that American economy needs several years of unemployment above 5% or 10% unemployment for one year to contain inflation.[71]  This proposal, if implemented, was sure to leave millions of Americans without jobs.

Cutting government programs is another prong of the austerity approach.  Both Ronald Reagan and Bill Clinton pursued the reduction of government spending by cutting federal programs, including welfare programs.  This policy has certainly had an adverse effect, causing pain and suffering among the most vulnerable members of society.  Summers rocked the Democratic establishment in 2021 by predicting that his party's excessive spending would cause inflation, insisting that government spending should be cut.[72]

The most obvious negative side to this approach is that it leads to much suffering for a great number of people who would have to endure long periods of unemployment and reduced government assistance.  Indeed, they would face enormous difficulties in trying to cope on their own with the deteriorating conditions of their life.  Predictably, this policy will inevitably lead to the growing gap between the rich and the poor.  In fact, this growing gap has been persistent in America and the world for the last several decades. The erosion of the standard of living for vast number of people is sure to generate social tensions and conflicts that will disrupt social peace and stability.  Although many proponents predict that this period of increased suffering will be limited, nobody can really tell how long it may last.  Few doubt, though, that the longer it lasts, the more unpredictable and dangerous will be the social and political consequences.

However, these adverse consequences do not exhaust all the negative effects of the austerity approach.  It may also profoundly distort the structure of the economy that will become more oriented toward the wealthy consumers, rather than the middle class.  Production will cater to those with money.  As a result, the market will offer more luxury real estate, more private jets and extravagant yachts, and more conspicuous consumption for the rich, rather than benefits for the middle and low ranks of society.  One should remember that mass production oriented toward the average citizen was what made American economy a success story.  An increased production of high-end goods is unlikely to replicate this success.

As many argue, inflation, even relatively small inflation, is dangerous for the economy.  Hrushikesh Mallick finds that "inflation rates have a significant adverse impact on economic growth."[73]  A number of researchers point out that the relationship between inflation and economic growth goes both ways.  While low productivity can result in inflationary tendencies, inflation also makes borrowing money more expensive, which slows economic growth.[74]  We must address inflation.  Yet on close examination, all proposed remedies augur ill for our economic future.  Many economists and economic planners have publicly voiced their pessimism.  They do not see that any of these remedies promise relief in the current conditions.  Using combinations of these remedies—as, for example, the Fed has been doing in the current inflation, by pursuing both policies of low and high interest rates--looks also extremely problematic.  As Fabio Vighi aptly summarized, "For most of us, then, the future seems to offer a choice between structural stagflation (stagnant economy with high inflation) and an abrupt deflationary depression--like a choice between bleeding to death and suffering a heart attack."[75]

## Understanding Inflation

*Inflation and Business Cycle*

As has already been pointed out earlier, there is no unanimity among economists and economic planners in their views of inflation.[76]  The only general conclusion that one can draw from their discussions is that inflation is a result of some complex imbalance in the economy; and that inflation is only one of the symptoms of this imbalance.  Researchers have failed to explain why economy becomes periodically unstable.  They merely recognize the existence of such periodic phenomena that they call "business cycle."

Business cycle is the pattern of economic booms and busts that are experienced by all developed economies.[77]  *Columbia Electronic Encyclopedia* attributes the first formulation of the theory of business cycles to French physician Clement Juglar who was the first to recognize, in 1862, that economic fluctuations associated with the boom-and-bust were a characteristic feature of all economic systems.[78]  The Great Depression that struck the United States and the world in 1929 was an important catalyst that stimulated much interest in business cycles—these periods of rapid economic expansions followed by economic slowdowns and contractions.[79]

Theories explaining business cycles are numerous and diverse.  They focus on different factors that supposedly trigger economic contractions in business cycles, but that hardly amounts to an explanation as to why business cycles are there in the first place.  Despite the fact that business cycles have been the subject of intense scrutiny for a very long time, the universal recognition today is that we simply do not know the why of business cycles.  The best that researchers do is merely point out the existence of economic fluctuations that they explain with such vagaries as "chaotic market processes."  Matthew Shapiro concludes that most theories "take the answer to this question to be axiomatic"—that is, the cycles are merely assumed to be part of economic reality.[80]  James Mirrlees echoes a similar view that "recessions are in some degree inevitable" and "are bound to happen.[81]  In his colorful metaphorical description Mirrlees compares attempts to steer economy away from downturns with efforts "to sail a straight line in a boat with wind direction constantly shifting, and sometimes blowing a gale."[82]

Neo-classical economists are very vocal and consistent in pointing to monetary interventions as perhaps the most significant factor that contributes to economic contractions.  Murray Rothbard is one of many who confidently claims that the "'boom-bust' cycle is generated by monetary intervention in the market, specifically bank credit expansion to business."[83]  However, even he recognizes that the problem of business cycles is "one of general boom and depression."[84]  He dismisses economic fluctuations as the source of depressions.  Rothbard writes:

> We may, therefore, expect specific business fluctuations all the time. There is no need for any special "cycle theory" to account for them. There is nothing here to account for a general business depression—a phenomenon of the true "business cycle" . . . The explanation of depressions, then, will not be found by referring to specific or even general business fluctuations per se.[85]

Rothbard finds that there is something unexplainable in the behavior of many experienced business people who are "misled" by the availability of cheap credit.  One can see Rothbard's sense of profound amazement at something incomprehensible when he writes:  "In short, how did all the country's astute businessmen come to make such errors together, and why were they all suddenly revealed at this particular time?  This is the great problem of cycle theory."[86]

In his reflection on the Great Depression, Milton Friedman, an acknowledged doyenne of neo-classical economics, also makes a claim that "monetary developments [in the early 1930s] were the major explanation for the depth and the length of the contraction."  He further explains:  "As I've said over and over again, I'm not saying that that [monetary developments] caused the initial recession . . . And I don't doubt for a moment that the collapse of the stock market in 1929 played a role in the initial recession."[87]

Thus, most neo-classical economists advise to take business cycles and depressions as a given and even embrace them.  Arguing that business cycles, including depressions, are an essential part of economy, Rothbard, for example, suggests that rather than fight depressions, we should change our perspective on them from pessimistic to optimistic.  We should view them as actually serving a useful purpose.  Downturns, in his view, are the way that economy "adjusts to the wastes and errors of the boom, and reestablishes efficient service of consumer desires."[88]  "The depression," Rothbard rhapsodizes, "far from being an evil scourge, is the necessary and beneficial return of the economy to normal after the distortions imposed by the boom.  The boom, then, requires a 'bust.'"[89]  Rothbard is not alone in proposing a change in attitude toward economic ups and downs.  According to James Mirrlees,  "The general conclusion [among economists today] is that we should encourage people not to worry too much about asset price fluctuations.  Then they will be happier; and they will not be so likely to reduce their consumption spending when markets crash."[90]

Not everyone agrees with neo-classical economists in viewing business cycles as an intrinsic part of economy.  Such 20th-century theorists as John Maurice Clark and Joseph Schumpeter have attempted to find cures for economic instability, rather than describe it, in the manner of many 19th century theorists, simply as a natural phenomenon.[91]  Tejvan Pettinger, among many, considers this view to be controversial.  Critics disagree with the view that economic downturns have a beneficial role to play because they "shake up" economy, weed out "inefficient" firms, and create incentives for cutting costs and operating efficiently.  They argue that in a recession, even "'good efficient' firms can go out of business leading to a permanent loss of productive efficiency."[92] Rendig Fels is another critic who rejects the notion of the inherent nature of business cycles. He sees "little evidence of a built-in tendency of the American economy to generate cycles."[93] Yet even detractors offer no insight as to why these cycles exist.

Indeed, there is a great deal of truth in the argument of neo-classical economists that monetary interventions are a major contributor to economic downturns in business cycles. However, one should recognize that the introduction of monetary interventions by the government was not a whim. It was largely a response to the existence of business cycles. The current inflation did not start merely because some policy makers decided to institute low interest rates. For example, one important reason for lowering interest rates to almost zero in the current inflation period was to prevent deflation and production decline caused by the global financial crisis that started in 2007. After the outbreak of the crisis the US Federal Reserve and other central banks around the world kept interest rates low for a prolonged period of time and have instituted other monetary policies to ensure that financial systems have plenty of liquidity.[94]

When John Maynard Keynes published *The General Theory of Employment, Interest, and Money* in 1936, the world economy was in ruins. The book was an important watershed in macroeconomics. Keynes's theory did not explain business cycles but it argued that monetary interventions by the government are the way to mediate their most adverse consequences.[95]

Initially, the response to Keynes's theory was limited to scholarly circles. It was only after WWII that the theory swept away the influence of the classical orthodoxy and became the main tool for guiding economies. The "Keynesian Revolution" got under way.[96] The most important factor in this new development was the rising wave of recessions in the 1950s and 1960s. During that period many economic planners and policy makers came to believe that there was a direct trade-off between unemployment and inflation. They came on the side of inflation to keep unemployment down.[97] Paul Samuelson and Robert Slow, two Nobel laureates in economics, forcefully argued in support of maintaining the price index at 4 to 5 per cent a year as "the necessary cost" of keeping employment around 3%.[98] When Nixon was blamed for the on-going recession in 1971, he reportedly quipped: "We'll take inflation if necessary, but we can't take unemployment."[99]

*The Nature of Production*

The connection often made between inflation and business cycle indicates that both these phenomena are interrelated. Inflation is one of the possible ways in which economic instability manifests itself; and so does downturn of production. In other words, inflation is associated with economic instability that is part of the business cycle. We do not know what causes this instability, nor do we understand why business cycles are there. What we do know, however, is that these phenomena are intimately related to production. Therefore, in order to understand inflation and business cycles we must look closer at the process of production.

We commonly view production in economic terms, as the action of making or manufacturing from components or raw materials, or the process of being so manufactured. According to one definition,

> Production is the process of making or manufacturing goods and products from raw materials or components. In other words, production takes inputs and uses them to create an output which is fit for consumption—a good or product which has value to an end-user or customer.[100]

In a contribution to *Economic Discussions* Sanket Suman defines production as "the organised activity of transforming resources into finished products in the form of goods and services; the objective of production is to satisfy the demand for such transformed resources."[101] Finally, Wikipedia offers the following extensive formulation:

> Production is the process of combining various inputs, both material (such as metal, wood, glass, or plastics) and immaterial (such as plans, or knowledge) in order to create output. Ideally this output will be a good or service which has value and contributes to the utility of individuals.[1] The area of economics that focuses on production is called production theory, and it is closely related to the consumption (or consumer) theory of economics.[102]

There is, however, another and broader view of production as a totality of human interactions with reality. German idealists tended to have such comprehensive view of production. In this thinking, cognition, or knowledge acquisition, is also a form of production. As Hegel put it, reality in this sense is "process and result rolled into one."[103] Karl Marx also viewed production in terms of the broad interrelationship between humans and nature. Consider the following passage from his *Economic and Philosophic Manuscripts of 1844* in which Marx writes: "Each of his human relations to the world – seeing, hearing, smelling, tasting, feeling, thinking, observing, experiencing, wanting, acting, loving – in short, all the organs of his individual being . . . are . . . in their orientation to the object the appropriation of human reality."[104] This view of production does not contradict the purely economic perspective; it simply represents a broader approach. It is in this broader approach that I propose to explore production and its place in human life.

Production is a form of interaction between humans and reality. It is a sensuous, or physical form of interaction. Humans are a product of the evolution. Therefore, human interactions with reality are also a product of the evolution. The evolution is a universal process that sustains our universe. Consequently, production is a result of the process that plays the most essential role in the existence of our universe.

The evolution originates in conservation that is ubiquitous throughout the universe. The roots of conservation are in the very unique nature of our universe. The universe is all that is there. Nothing can come into it from outside because there is no outside; nothing can disappear from it because there is nowhere to disappear. Consequently, everything must be conserved.

Conservation requires resources; and resources are always limited. Therefore, new resources are vital for conserving our universe. These necessary resources cannot come into our universe from outside. They have to be created inside the universe. Therefore, conservation requires creation of new and increasingly more powerful levels of organization that offer new possibilities that have not existed prior to their creation. These new possibilities offer access to new resources; they represent such new resources. Thus, the universe is impossible without the creation of new and increasingly more powerful levels of organization;[105] and the creation of such levels of organization, or their production, is what the evolution is all about.

Since humans are products of the evolution, their interactions with nature have inherited the main features of the process of creation. We also live under the constraint of limited resources. Therefore, in order to sustain our life we have to create new and increasingly more powerful levels of organization that provide access to new resources. This is the essence of our production, including economic production.

*Value Creation and Inflation*

Newly created and increasingly more powerful levels of organization offer new possibilities that are the main resource for sustaining our existence. Therefore, these new and increasingly more powerful levels of organization represent an enormous value for us since they sustain our existence. They are the source of value. Value creation is the main function of our interactions with nature; it is also the main function of our economic activity, or production.

What does the creation of new and increasingly more powerful levels of organization involve? A level of organization consists of interconnected functional operations. In order for these operations to persist, they must be conserved. Action is what conserves operations: the more operation is active, the better it is conserved.

In order to bring an operation into an active state, or to activate it, the operation needs something that will trigger its activity. The more often the operation is triggered into action, the more active it is, and the better it is conserved. Consequently, the more triggers an operation has, the better it is conserved.

If an operation combines with another operation to create a new whole, it will double the number of triggers that bring it into action—one that triggers one operation and another that triggers the other operation. Since interactions among functional operations conserve them, such interactions are favored by the evolution. By combining with each other, operations acquire new possibilities and, therefore, become more powerful in comparison with when they were on their own. Each new level of organization that we create, or produce, increases our power and, thus, represents value. In order to conserve this value, we must create another and even more powerful level of organization, or a new value. Constant value creation is the main function of production.

Thus, interactions among functional operations create new levels of organization that offer more possibilities and, consequently, are more powerful than those from which they have emerged. This added power represents value. Production is about the creation of value, which means that production involves the creation of new and increasingly more powerful levels of organization. The only way to conserve value is to conserve the level of organization that represents this value; and the only way to conserve this level of organization is by creating a new level that is more powerful, and consequently has more value, than the one from which it has emerged.

The conclusion that follows from the above is that the primary function of production is to conserve functional operations sustained by a particular level of organization. Production is ultimately about conservation; and conservation can only result from creating new values. Failure to create new and more powerful levels of organization means that existing possibilities have not been realized. Without realizing these possibilities, they cannot be conserved; in other words, value is not conserved. Non-conservation of value is a sure sign that production is not efficient and that available resource are either underutilized or simply wasted. Inefficient production does not create value and, consequently, sustains losses. Production losses are at the root of economic instabilities and disruptions.

All economic values have monetary equivalents. Money plays the most essential role of realizing the possibilities that each value offers. We use money to attract resources and labor to production. Money does not have intrinsic value, after all money is a mere convention, a piece of paper with something printed on it--and now even not that; today money is merely an electronic signal transmitted from one source to another. The value of money is the equivalent of the real value. When real value is not conserved, when all possibilities associated with this value are not realized, the real value these possibilities represent diminishes, or depreciates; and when value depreciates, the money equivalent of this real value that is supposed to be used for realizing these possibilities also loses its worth. Money becomes cheap and the cheapening of money results in price increases, which is the essence of inflation.

**The New Economic Practice**

*The Model for the New Economic Practice*

The philosophic exploration of production shows the connection between production as an evolutionary phenomenon and the universal process of creation. This exploration leads to several important conclusions. First of all, it makes clear that inflation is an aberration that is in no way intrinsic to production. There is nothing in the process of creation that warrants this aberration. The growth in the supply of money is not the ultimate source of inflation. It is merely an effect of the depreciation of money that results from inefficient production that leads to the depreciation of real value and the monetary equivalent of this real value. The depreciation of real value indicates that our production fails to conserve this value.

Production must be efficient. There is hardly a single economist, businessman, economic policy maker, and even a lay individual who would dispute this maxim. Efficiency results only from the full utilization of all available resources and possibilities. Wasted resources and squandered possibilities make economy inefficient. Efficiency is what makes economy possible; it is the pillar on which our economies rest. All negative economic phenomena—instabilities, crises, inflation, unemployment, and others—are results of inefficient production. In order to understand our economic problems, including inflation, we must understand why we fail to use resources efficiently, create wastage; in other words, why we do not conserve value.

The philosophic investigation that traces production to the universal process of creation shows that production is an evolutionary phenomenon. It emerges in the course of the evolution driven by the universal process of creation and, therefore, production inherits all the main features of this process. The distinct feature of the process of creation is the full utilization of all available possibilities. In fact, this process can only work on the basis of such full utilization. Anything less than full utilization makes the creation of new and increasingly more powerful levels of organization impossible. Thus, full efficiency is the essential property of the process of creation. Therefore, the main requirement of efficient production is the full utilization of all available resources.

There are other important features of the process of creation that efficient production must replicate. One important feature of the process of creation is universal inclusion. The full utilization of all available possibilities is the key to the creation of a new and more powerful level of organization. Only such utilization conserves all available possibilities, which is the only way to create a new level of organization that is more powerful than the one from which it has emerged. Only the combination of all available possibilities makes the new emerging level greater, or more powerful, than the sum total of its parts. If any possibility is excluded, it will not be conserved; in other words, it will be wasted, which is how we define inefficiency. As a result of exclusion, the emergent level of organization will not exceed the power the level from which it has emerged.

The process of creation is a dynamic process. It constantly evolves. The feature that makes such dynamism possible is the balance between equilibrium and disequilibrium that one can observe in the process of creation. The equilibration of specific operations (i.e., possibilities) increases equilibrium. However, equilibration also creates a new and more powerful level of organization. Since this new level has greater power, it represents disequilibrium. Thus, as equilibrium grows, so does disequilibrium. The two are always in balance and complement each other.

The creation of combinations conserves operations (possibilities). They do not disappear. No operation consumes another operation. Both are conserved. No operation that belongs to a given level of organization is superior to another operation sustained by the same level. Therefore, their interactions are interactions among equals, which means that these interactions are non-hierarchical. However, in the course of interactions, operations create a new and more powerful level of organization. Therefore, they create a hierarchy. Consequently, the balance between hierarchical and non-hierarchical interactions is another important feature of the process of creation. Non-hierarchical interactions create new levels of organization and hierarchical interactions conserve and optimize these new levels.

The above features are those that have already been recognized as having the essential role in the process of creation; they make it possible. In order to create an economic practice that would be able to replicate the process of creation in its efficiency, all these features must become part of the new economic practice. There may be some other features that we still do not know about and that we will recognize as we continue to learn more about the process of creation in the course of the realization of the new practice.

*Changes Required for the New Economic Practice*

In order to make our economy efficient, avoid wastage, and prevent instabilities, the new economic practice should use the process of creation as its main organizing principle. In this case, the new economic practice will include all the main features of the process of creation. As a result, the new economic practice will be different from the current practice in some critical respects. This new practice will also lead to rethinking of some key concepts that are relevant to economic practice.

In contrast to the existing practice, the new practice will involve a much greater emphasis on creativity. Investing in the production of new ideas is clearly an emerging trend in the current practice. In the new practice, investments in the production of new ideas will constitute the increasingly growing portion of total investments. Moreover, much of these new investments will be focused on radical innovation that is associated with the emergence of new and increasingly more powerful levels of organization.

1. The New Conception of Management and Leadership

The emphasis on creativity will inevitably affect our conception of management and leadership. In contrast to the current economic practice that is dominated by hierarchical interactions, there will be a much greater emphasis on the role of non-hierarchical interactions. Non-hierarchical interactions play the main role in creating new combinations. They are the source of creativity. Hierarchical interactions conserve and optimize what non-hierarchical interactions create. As has been explained, the process of creation requires a balance between the two types of interactions. Maintaining this balance will be one of the main preoccupations of the new economic practice.

The top-down approach dominates our current economic (and not only economic) practice. Indeed, there are some experimental trends that try to moderate hierarchical domination. But even these new trends do not go nearly far enough in balancing hierarchical and non-hierarchical interactions. For one thing, these innovative trends limit the scope of balancing only to top economic and managerial elites and excludes a large number of people involved in the process of production and exchange, including but not limited to workers, employees and even small and medium-size businesses. In the United States, for example, small and medium size businesses do not qualify for a generous support of the kind that goes to corporate giants, such as GM, Ford, or major mega-banks. While the market certainly has a non-hierarchical structure, our managerial culture remains by and large hierarchical.[106] The top economic and managerial elites essentially adhere to principles of hierarchical control, rather than to the non-hierarchical mode associated with the market. The dominant neo-liberal approach in economic management has been pursuing and to a significant degree has achieved the merger of the government and economic elites. Rather than combine the hierarchical principle of government bureaucracy with the non-hierarchical principle of the market, this approach has strengthened the concentration of power in the elites and enhanced the hierarchical principle in our society. It has not balanced hierarchical and non-hierarchical interactions.

There is a growing number of researchers who recognize the need for a genuine combination of hierarchical and non-hierarchical principles. One popular trend is the so-called hybrid solutions, that is, solutions that still see hierarchical and nonhierarchical interactions as ontologically separate but seek some format in which their coexistence and limited cooperation can become possible. These solutions are largely eclectic and do not achieve a true integration.[107] John Kotter, the chief innovation officer at Kotter International and a professor emeritus of the Harvard Business School, typifies this approach. In his view, hierarchies and networks are two separate structures that excel at what they do best. Kotter recognizes that hierarchies are very good at optimizing and are capable of effecting small and medium-size changes but cannot perform large-scale innovative transformations. He explains:

> But I am referring to something far bigger: large-scale organizational change, such as a company redesigning its entire business model, or accomplishing its most important strategic objectives of the decade, or changing its portfolio of product offerings. *And there is no evidence to suggest that the Hierarchy allows for such changes, let alone that it effectively facilitates them.*[108]

In Kotter's view, the future lies in the coexistence of the two structures in one business organization. In his own words:

> All of this has led me to believe that the successful organization of the future will have two organizational structures: a Hierarchy, and a more teaming, egalitarian, and adaptive Network. Both are designed and purposive. While the Hierarchy *is as important as it has always been for optimizing work, the Network is where big change happens.* It allows a company to more easily spot big opportunities and then change itself to grab them.[109]

Coexistence is a far cry from a genuine integration. It presupposes that the competition between co-existing entities will continue and will be merely moderated. Ultimately, they will not cooperate but they will try to avoid interfering with each other. This solution is certainly not enough. The two types of interactions should not merely co-exist, which is a very unstable arrangement based on constant competition, but truly cooperate and complement each other. They should be part of the same decision-making process, not merely the two sides in a compromise decision. Compromise solutions involve emphasizing commonalities and suppressing differences. Yet, differences, not commonalities, are the main source of radical innovation.

Hybrid solutions offer a rich plethora of interesting ideas regarding possible mechanisms of interactions between hierarchies and networks. However, as all eclectic solutions, they do not have a solid theoretically grounding and tend to have internal contradictions. Nothing illustrates this shortcoming better than the discussion of such critical subject as the relationship between leaders/managers and networks/employees. Opinions on this score vary widely: from a more activist role of leaders/ managers as enablers[110] to a weaker role of regulators and filterers of external information,[111] to an even weaker role as facilitators of critical discourse and enhancers of local interactions among network agents.[112] Some even believe that the desired goal can be achieved without structural changes by merely modifying the rationale for the role of hierarchies and by educating managers in the values and merits of organizational democracy. Martin Clarke and David Butcher, for example, see education and the principle of voluntarism they borrow from political philosophy as vehicles for reconciling hierarchies and networks in organizational structures.[113]

There is no doubt that the literature on hybrid solutions certainly deserves serious attention. It addresses many aspects of what is obviously a very complex and seemingly intractable problem. Many of ideas articulated in hybrid solutions are undoubtedly very useful. But even all together, they hardly measure up to the magnitude of the task, which leaves quite a few researchers dissatisfied and vying for a comprehensive solution. In their essay "Simplistic vs. Complex Organization: Markets, Hierarchies, and Networks in an Organizational Triangle," Wolfram Elsner, Gero Hocker and Henning Schwardt make an argument for just such a comprehensive solution. In their view, "… pure market and hierarchy, including their potential formal hybrids, are an empirically void set." Rather, "coordination forms" in the real world, they argue, "have to be conceptualized in a fundamentally different way. A relevant organizational space must reflect the dimensions of a complex world."[114]

In making their appeal to complexity of the real world, Elsner, Hocker and Schwardt suggest that the division between hierarchical and non-hierarchical interactions is not real, it is merely conceptual;[115] that in reality, the two types of interactions are closely entangled with each other, although they fail to explain the nature of this entanglement. Numerous other researchers support the approach that centers on the entanglement of hierarchical and non-hierarchical interactions and the complexity of their relationship. Antoine Danchin points to the ubiquity of networks and hierarchies in nature and their complementary relationship.[116] Joan Roelofs challenges the simplistic view of networks as spontaneously resistant to hierarchies and naturally prone to democracy. As she maintains,

> …some participants in network governance are vastly more powerful than others. As for "civil society" organizations, support from corporate or private foundations is essential to almost all civil rights, social justice or environmental organizations that wish to be viable and visible; the funders exert control in many ways.[117]

While the above perspectives serve as a valuable source of insights, they ultimately do not resolve the problem of the relationship between networks and hierarchies. Despite their astute and nuanced observations on the nature of this relationship, they still see hierarchies and networks as ontologically separate. In their view, tensions between networks[118] and hierarchies can only be ameliorated, but they will ultimately always remain and be a potential source of conflict.[119]

The perception that networks and hierarchies are polar opposites, perennially in tension and conflict with each other, contradicts what we know about systems in nature. As has been explained earlier, systems conserve themselves by forming bonds, or what Humberto Maturana and FranciscoVarela called structural coupling,[120] with other systems in their environment as part of the process known as self-organization, thus creating new organized totalities.[121] The process of creating a new organized totality gives rise to the operation that regulates the functioning of this totality. And that is what a system is: an organized totality of coordinated operations with a common regulatory mechanism.

Since the regulation of a system is a product of combining the capabilities of its constituent parts, the level of organization that supports regulation is more powerful than the level of organization of each subsystem or their sum total. The emergence of this more powerful level of organization creates a hierarchy. As one can see, regulation is a product of interactions among subsystems. It supervenes on local interactions and vitally depends on them for its own existence.

The regulatory function also needs to be conserved. For this reason, it has to form strong bonds that would activate it; and first and foremost, it should have strong bonds with the subsystems it regulates. The process of forming bonds between the global level of regulation and the level of local interactions results in the integration of the two levels of the system and the adaptation of local interactions to the global operations. Regulation can facilitate such adaptation. When the weaker operations adapt to the more powerful level that sustains regulation, they change and gain in power; the re-equilibration of these enhanced operations increases the power of the regulatory level. Thus, the entire system evolves.

This observation regarding systems in general suggests that in social systems the role of leaders and hierarchies, which also operate at a more powerful level of organization, must be very similar. By virtue of their position, leaders can enormously facilitate the integration of systems because they have access and can observe both the global and the local level of interactions. In order to integrate the system they regulate, leaders must resort to reflective coding—the procedure that Gödel used in his famous proof of consistency and completeness.[122] It is a creative task because it creates a level of organization that can incorporate both global functions and local interactions as its particular cases.

This role leaders and hierarchies has nothing to do with command and control, that is, transmitting decisions from those above to those below and overseeing their implementation. Leaders must appreciate the enormous creative power of local interactions and be closely attuned to their variations and modifications. Since they rely, or supervene, so much in what they do on interactions among network agents, or subsystems of the system, they should promote, regulate, and facilitate these interactions, not dominate them and impose on them their will. It is a sensitive, delicate, and highly creative role that involves both cooperation and two-way adaptation. Those who operate at the global level and those involved in local interactions are, in a way, equal participants in a common creative enterprise of ensuring the conservation and evolution of the system that they constitute.

Because of their location in the liminal space between the system and its environment, hierarchies and leaders are in a position to reflect critically (that is, observing at the same time the system and also themselves in performing their function)[123] on all interactions among all the local agents and subsystems of the system. The latter, by virtue of their position, can reflect only on local interactions.  For this reason, the position of leaders makes possible for them to see new and more powerful possibilities emerging in interactions within the system, as well as recognize, promote, and facilitate the utilization of these possibilities.

The creation of new and increasingly more powerful levels of organization that propels the system's evolution is incompatible with the relationship of exclusion and domination.  It requires cooperation and close interaction in common creative work. Such cooperation can only be effective if there is a balance between hierarchical and non-hierarchical interactions, between hierarchies and networks--managers and leaders, on one hand, and employees, on the other.[124]  Leaders should not see their role as that of ultimate arbiters whose word is decisive and final—far from it.  The notion of a leader as the ultimate arbiter without whom there will be chaos and instability is a result of a profoundly flawed view that excludes the process of creation from its frame of vision and, as a consequence, leads to a failure in understanding how systems function and evolve.  This view makes impossible to have clear and rational validation criteria that can help choose the most powerful level of organization.  As has been argued elsewhere, the current approach largely relies on subjective choices of those at the top of the hierarchy.[125]  The lack of such objective and rational criteria of validation is the main reason why we now tend to defer decisions to top managers.  In the absence of such criteria, all decisions are subjective and all are equal.   Recognizing all decisions as equal is likely to lead to chaos and instability and nobody wants to argue for disorder.  As a result, the common current default is to defer to the decision of those who are at the highest level in the hierarchy because even a bad decision that preserves order is deemed better than chaos and instability.  How many businesses have paid a heavy price for limitations of those at the top of their hierarchies?

2.  Assessment of Economic Performance

The current economic practice uses value as the most important criterion for assessing results of production.  However, we greatly differ in our ways of understanding value.  Some use market share in determining value, others profitability, still others prices of company's shares on the stock market; and there is more.  The choice depends on how one defines value.

There are many definitions of value currently in use.  IGI Global, for example, identifies eight different definitions of value and its list is not the most extensive.[126]  With this number of definitions, their differences and diversity do not come as a surprise. Perhaps the most common formulation defines value as "a measure of the benefit provided by a good or service to an economic agent."  Currency is the most common measure of value; consequently, money is the most common way to quantify value. However, it is not the only measure of value.  In another formulation value is "the maximum amount of money a specific actor is willing and able to pay for the good or service."[127]  Caroline Benton, for example, defines value in terms of preferences of individuals who "determine the economic value of a good or service and the trade-offs that they will be willing to make to obtain it."[128]

Although in the above formulations money usually objectifies value, the influence of subjectivity in valuation is unmistakable: value is based on preferences. Preferences do not have to be monetary either. Yusuke Kuwayama, Justine Huetteman, and Bethany Mabee, for example, do not consider money as the only and even the most important way of assessing value. "Simply put," they argue, "things that have 'value' are useful to you, improve your situation, or simply make you happy or more secure."[129] All these are more or less subjective factors.

There are some formulations that try to see value as something calculable. The emphasis on calculability is supposed to give value the appearance of something objective. Yet even such formulations cannot escape subjectivity. *Market Business News* defines value as "a calculation of the profits an asset has either produced or may produce in the future." Value is "a measure of the benefit a product or service provides an economic agent (person or company)." Ultimately, the willingness of the agent to pay for something, his or her preferences, still appear in the estimation of value. Finally, all formulations stress that economic value is not the same as market price that is regarded as more objective than individual preferences.[130]

Finally, there is this gem of a definition that makes the determination of value a total exercise in futility. A team from the Corporate Finance Institute offers the following formulation:

> The economic value of a business is the business's contribution to the global gross domestic product (GDP). The most common method of estimating economic value is the counter-factual method. The counter-factual method states that the economic value of a business is the difference between the current global GDP and the hypothetical global GDP if the business did not exist.[131]

Even a brief overview of some of the examples of the current definitions of value shows their inadequacy. They are not uniform. They often contradict each other. Also, and probably even more importantly, these definitions are to greater or lesser extent subjective. Basing our assessment of economic efficiency on such subjective understanding of value cannot result in an objective valuation. We need a uniform and objective understanding of what constitutes value. The economic practice that uses the process of creation as its main organizing principle offers such understanding.

As the earlier discussion of the process of creation makes clear, efficiency depends on the use of resources and possibilities. A system—any system, including economic ones—is efficient if and only if it fully utilizes all resources and possibilities available to it. Only by combining all these resources and possibilities, a system can create a new level of organization that will make it more powerful. Such system will conserve itself and will make an evolutionary advance. The new level of organization will offer access to new resources and new possibilities.

In order to be efficient, a system must be inclusive. In order to survive, business organizations must know their resources, both physical and mental. The latter are particularly important in our time because of the growing emphasis on creativity and the production of ideas. If resources are known, one can easily determine whether production uses all of them or just some. Also, production involves the creation of combinations of all available operations. In other words, each operation should establish connections with all other operations. If a system that has n-number of operations functions efficiently, it will produce the number of operations that will be equal to $n^2$. In other words, an efficiently functioning business organization will have an exponential growth.[132] Conversely, if a system demonstrates an exponential growth, it must be functioning with full efficiency. Profits will necessarily reflect this efficiency.

Inclusivity and exponential growth are not subjective criteria. They do not depend on individual preferences. These criteria are derivatives from the process of creation that is not our subjective construct. This process is not our creation. On the contrary, this process has created the human race. This process is the source of value that is determined in objective terms of inclusivity and exponential growth.

3. The new conception of production and consumption

The preceding discussion of the new economic practice focused primarily on production. However, there is another dimension that is also relevant for determining economic efficiency. This dimension is consumption. Efficient economy is one in which all products are consumed and nothing goes to waste. Economic efficiency includes both production and consumption.

Despite the fact that efficient economy involves a close interrelationship between production and consumption, there is a strong tendency in our economic science to separate the two, even though it recognizes, at least to some extent, their close interrelationship. Our economists and business people accept the notion that ultimately we should strive for a full utilization of what we produce. Otherwise, our production will be inefficient and sustain losses and waste.

However, despite this recognition, our economic thinking also sees production and consumption as two ontological opposites. This thinking generally associates production with appreciation, or value creation, and consumption with depreciation or the erosion of value. We know very well, for example, that when we buy a new car, the car will lose a significant part of what we paid for it as soon as it leaves the parking lot of the dealership.

The analysis of the process of creation shows that production and consumption are two closely interrelated aspects of one integral whole. They are merely analytical categories, not ontological entities that exist separately from each other. When subsystems form bonds with each other, they assimilate each other. In other words, they include each other in their own functional operations, which is essentially a form of consumption. However, the inclusion of a different entity into the subsystem's functional operations modifies these operations, thus producing something new that did not exist prior to inclusion. Moreover, even to be open to a possibility of inclusion, a subsystem must create a construct that allows it to perceive another subsystem in its environment. Such perception does not come automatically. A system must have a capacity to perceive difference. Regulation has this capacity to perceive something different form the entity it regulates.

Our perception of reality offers a good illustration of this point. In order to perceive, our mind has to create a mental construct that makes this perception possible. The act of perceiving assimilates the perceived object and thus stabilizes and conserves the mental construct that makes the perception of this particular object possible. Thus, the production of perceived reality goes hand in hand with consumption of this reality. We cannot perceive what we have not first constructed. Infant's experience of reality is defined by inherent sensory-motor operations. Mental constructs that make possible for the infant to perceive permanent objects are a result of the combination of sensory-motor operations that create permanent mental images.[133]

Thus, production and consumption go hand-in-hand together. They are both aspects of the process that creates new properties and new possibilities. Production and consumption are analytical, rather than ontological categories. As the analysis of the process of creation shows, the conservation of a given level of organization creates a new and more powerful level of organization. The emerging level of organization supervenes on the level from which it emerges; in other words, the level of organization that gives rise the new level of organization is a resource that is consumed in the process that creates the new level. One can also represent this relationship between consumption and production as a balance between equilibration and the production of disequilibrium. In this conception, disequilibrium is a resource for equilibration; and equilibration results in a new disequilibrium.

What practical consequences will the recognition of the unity of consumption and production? How will it affect our economic practice?

Organizing our economic activities around the process of creation will end the tendency to dissociate production from consumption; it will make the interrelationship between production and consumption effective and efficient. By complementing each other, they both will be able to grow exponentially and will make possible an exponential growth of our entire economy. One can see the contours of this new economic organization in the comment of Alan Webber who concludes: "In the end, the location of the new economy is not in the technology, be it the microchip or the global telecommunications network. It is in the human mind."[134]

In contrast to ordinary goods, knowledge does not depreciate when used. On the contrary, it appreciates. In other words, its value grows. Our assimilation of ideas creates new and increasingly more powerful levels of organization that give rise to even more innovative ideas, approaches, and decisions. By producing new and increasingly more powerful levels of organization we generate new knowledge and ideas that lead to economic growth. Knowledge, for example, is one important product that does not depreciate. It only appreciates when consumed, as its consumption leads to new and increasingly more powerful levels of organization that give rise to new knowledge and ideas. As Thomas Davenport and Lawrence Prusak have noted, "ideas breed new ideas, and shared knowledge stays with the giver while it enriches the receiver."[135]

The realization that production and consumption are intimately related will change our patterns of investment. The economic practice that uses the process of creation as its main organizing principle will focus investments on the creation of new and increasingly more powerful levels of organization, not merely on putting new items on store shelves. Such new pattern of investment will be becoming increasingly important with the on-going shift of the emphasis in our economy from production of things to production of knowledge and ideas.

The change in the pattern of investment will enhance our production and economic efficiency. In today's economy many goods that reach the market often face no demand. They are either drastically discounted or completely trashed. With the new pattern of investment and the emphasis on the production of knowledge and ideas we will avoid this waste. As our production grows exponentially, so will our consumption; yet we will not be facing a situation that what we are not able to consume what we have produced. Just like our production, our consumption can also grow infinitely; and this growth will not be posing any threats to our planet or the universe. On the contrary, they will help us in solving numerous problems that we face. Moreover, they will also help us anticipate future problems and find their solutions even before these problems emerge. Finally, the new patterns of production and consumption will help us live better and have a more satisfying and happier life.

4. Education

The importance of knowledge for economic production is hard to overestimate. Knowledge has always been a major contributor to economic progress. Its contribution is particularly important in this day and age when our economy puts premium on knowledge production and creativity; our economy is increasingly about knowledge production.[136]

Education is one important sphere of our civilization that directly relates to knowledge. Traditionally, storage and dissemination of the existing knowledge has been the main preoccupation of education. By transmitting knowledge to new generations of young men and women, the system of education prepared them for becoming productive members of society. Modernity self-consciously used education to set our civilization onto a new course. Since the very beginning of the industrialization, the connection between education and economic progress acquired particular importance. Educational institutions became the main breeding ground where students were trained to become part of the labor force that propelled the industrial development. They became the workers, technicians, managers, business organizers who built and operated the growing number of factories and plants. The connection between education and economic progress has become particularly important in this day and age when the production of innovative ideas, approaches, and decisions has become the most important production factor.

Due to the close association between education and economic progress, economic needs have always shaped our educational system. There is no doubt that the transition to the new economic practice will necessitate changes in our education and teaching methodology.

As has already been pointed out, educational institutions were traditionally preoccupied mostly with the dissemination of knowledge. Our educational system exposed students to the vast body of information and skills that were accumulated in the course of the evolution of human civilization. However, it rarely, if at all, taught students the habits and skills required for producing new knowledge. Indeed, many talented young men and women engaged in knowledge production and became the source of radical innovations. However, their success owed very little to the educational system. The manifestation of their genius occurred contrary to this system, rather than because of it.

Our educational system does pitifully little to enhance and develop the creative capacity of young people. Generally, our educational institutions relegate creative type of activities to art and literature courses.[137] There is very little room for creativity in mainstream courses in sciences, including social sciences, and math. As a result, more often than not, education stifles students' creative impulses and suppresses their capacity to create. This situation is hardly an accident. Our knowledge of what makes creation possible is meager. The sad fact has been and remains that our understanding of the process of creation remains very rudimentary. As a result, our ability to control this process is extremely limited. Margaret Boden, one of the pre-eminent researchers in the field, draws the following conclusion in her influential book on creativity:

> Our ignorance of our own creativity is very great. We are not aware of all the structural constraints involved in particular domains, still less of the ways in which they can be creatively transformed. We use creative heuristics, but know very little about what they are or how they work. If we do have any sense of these matters, it is very likely tacit rather than explicit: many people can be surprised by a novel harmony, but relatively few can explicitly predict even a plagal cadence.[138]

This situation must change with the onset of the new economic practice. Some radical innovations must take place in our theory of education and teaching methodology.

It is beyond the scope of this study to provide a detailed description of the kind of changes that are in order. The transformation of our education is a project that is only beginning. Many of the required changes will become clear only in the course of this project's implementation. However, the description of some essential features is possible today.

Whatever new forms our education will take, the process of creation should be their main model and inspiration. Therefore, the recognition of the importance of this process, further study of this process, and a comprehensive understanding of it are essential. Our educational system should use the process of creation as its main organizing principle.

The new teaching methodology must include important features that are known and characterize the process of creation. Universal inclusion and empowerment must constitute the basis of the new methodology. Teaching should also observe important balances that sustain the process of creation: the balance between equilibration and the production of disequilibrium, or the balance between equilibrium and disequilibrium, and the balance between hierarchical and non-hierarchical interactions. One can find a more detailed description of the proposed innovations elsewhere.[139]

**Conclusion**

Economic production is, in more ways than one, a very important building block of our civilization. It provides means for our sustenance, helps us control our environment, and makes our life more comfortable, enjoyable, and satisfying. But its importance transcends these utilitarian purposes. Economic production strengthens and develops our capacity to create, reinforcing our connection to the universal process of creation and, thus, to our entire universe.

This article argues that our economic problems, of which inflation is only one, are due to the fact that our economic production is inefficient. Its inefficiency is a result of the inadequate use of resources and possibilities available to us, most importantly human capacity to create an infinite number of new and increasingly more powerful levels of organization. This inadequacy is not fortuitous; it is not a result of our mistakes or lack of industriousness and diligence. It is a result of our lack of understanding of the roots of our own existence—the universal process that led to the rise of humanity and has been driving the evolution of human civilization. To this day, we have not grasped the important role this process plays in our relationship with reality.[140]

The article shows the connection between economic production and the process of creation. It also shows that the reason why our economy is inefficient is the fact that we do not know what factors makes the process of creation so efficient; consequently, we cannot replicate these factors in our economic activities. As the article shows, these factors include universal inclusion, the balance between hierarchical and non-hierarchical interactions, as well as the balance between equilibrium and disequilibrium. Only full utilization of all available resources and possibilities can create new and increasingly more powerful levels of organization that can give rise to new ideas, decisions, and approaches. Such efficient use of resources will lead to constant, stable, and exponential economic growth without inflations, business cycles, and economic contractions and crises.

Understanding the process of creation is only the first step in addressing the problem of our economic inefficiency. As this article has argued, we have to acquire knowledge of the process of creation and use this knowledge in establishing the new economic practice that will use the process of creation as its main organizing principle. Such new practice will involve changes in investment patterns, conception of management and leadership, as well as re-conceptualization of consumption and the relationship between production and consumption.

Indeed, some may make an argument that even though we do not understand the process of creation and its role in our relationship with reality, our civilization has been able to make a remarkable progress in the course of its history. This fact only reinforces the main point made in this article. It shows the enormous power of the process of creation that has helped our civilization to evolve to this point despite our disregard of this process and our failure to utilize fully its infinite possibilities. In a way, we have been able to make a remarkable progress despite ourselves.

However, our civilization can progress only so far if it continues to ignore the process of creation. The way we have pursued progress so far has limitations.[141] In order to continue the progress of our civilization into the future, we need to embrace, understand, and use this process efficiently to transcend our self-imposed limit. Our production has reached a limit of what it can achieve without availing ourselves fully of the enormous resources and possibilities that the process of creation offers. Our economic production has evolved to the point when our capacity to create is rapidly becoming the most important factor without which our production simply cannot evolve any further and help us in solving the problems we now face.

Our civilization badly needs new ideas,[142] including new ideas in the way we run our economies. So far, there have been two main approaches in organizing our production. Liberalism promotes one approach and socialism another. Both versions ignore the process of creation; and both have reached their limit. The many disruptions of the world economy are an eloquent indication of the inadequacy of the current approaches.

Capitalism is the term that has been much abused and maligned; the meaning of capitalism has been perverted and ideologically misconstrued. Capitalism is essentially about the growth of capital. Capitalists are people who focus their skills, energy, and resources on increasing capital, that is, on creating new value. The rise of capitalism represented a realization that making capital grow and creating value is so important that we should concentrate our will and resources on promoting constant growth. The growth of capital has become a self-conscious goal.

In this sense, Western liberalism and Marxist socialism are not very different from each other, despite their assertions to the contrary. Marx's teaching that has probably done more than anything else to discredit capitalism is in this sense very capitalist in its spirit and orientation. It is also about capital growth; it is also about value creation. In fact, Marx's main argument against capitalism is that capitalism has flaws that do not allow growing economy indefinitely. Socialism, according to Marx, was supposed to remove these fetters and to ensure unlimited growth. His argument for the capacity of socialism to accomplish this task has been a dismal failure. However, the passion and conviction that ring in this argument are genuine; they reflect his aspiration for infinite growth that is essentially "capitalist" in its nature in the sense in which capitalism has been defined above. The similarity in the pursuit of economic growth between liberalism and socialism is perhaps one reason why Western liberalism has proven to be receptive to socialist ideas. Many even mainstream progressive liberals in the United States today see solutions to the current problems in socialist policies, such as redistribution of wealth and the involvement of the state in managing the economy.

The rise of capitalism was an important evolutionary development. The idea that the survival of our civilization vitally depends on perennial creation of value and capital growth is our articulation of what we have inherited in the course of the evolution from the process of creation. Our anthropocentric bias has prevented us from grasping the cosmic significance of this process. We have failed in moving beyond our vague intuitions about it. The anthropocentric attribution of the emergence of capitalism exclusively to humans and their decisions has overshadowed the intuition of early capitalists who believed that their ability to increase capital was divine in its origin. These "secular monks" created much value and acquired enormous wealth, yet they did not use this wealth for the sake of hedonistic enjoyment; they relished with the abandon of religious ecstasy in their very capacity to create value that they felt brought them closer to God the Creator.